\documentclass{PoS}

\title{Singularities of QCD in the complex chemical potential plane}

\ShortTitle{Singularities of QCD in the complex chemical potential plane}

\author{\speaker{Shinji Ejiri}%
\thanks{This work has been authored under contract DE-AC02-98CH1-886
with the U.S. Department of Energy.}\\
Physics Department, 
Brookhaven National Laboratory,
Upton, New York 11973, USA\\
E-mail: \email{ejiri@quark.phy.bnl.gov}}

\author{Hiroshi Yoneyama\\
Department of Physics, Saga University, 
Saga 840-8502, Japan}

\abstract{
We study the thermodynamic singularities of QCD in the complex chemical potential plane by a numerical simulation of lattice QCD, and discuss a method to understand the nature of the QCD phase transition at finite density from the information of the singularities.
The existence of singular points at which the partition function $({\cal Z})$ vanishes is expected in the complex plane.
These are called Lee-Yang zeros or Fisher zeros.
We investigate the distribution of these singular points using the data obtained by a simulation of two-flavor QCD with p4-improved staggered quarks.
The convergence radius of a Taylor expansion of $\ln {\cal Z}$ in terms of the chemical potential is also discussed. 
}

\FullConference{The XXVII International Symposium on Lattice Field Theory\\
		 July 26-31 2009\\
		 Peking University, Beijing, China}

\begin{document}

\section{Introduction}
\label{sec:intoro}

The existence of singular points at which the partition function $({\cal Z})$ vanishes is expected in a complex parameter plane.
These are called Lee-Yang zeros or Fisher zeros \cite{LeeYang}.
The scaling analysis of such singularities in the complex parameter plane is an interesting approach to understand the nature of phase transitions. 
A few years ago, the thermodynamic singularities of QCD in the complex chemical potential $(\mu_q)$ plane were discussed by M. Stephanov using universality arguments in the vicinity of the critical point and a random matrix model \cite{steph06}.
It is interesting to perform numerical simulations of lattice QCD and to compare the results with the predictions from the universality arguments. 

Moreover, the study of the singularities in the complex plane is important to estimate the radius of convergence of a Taylor expansion in terms of $\mu_q$. 
The Taylor expansion method \cite{BS05} is widely used for the study of the equation of state at finite $\mu_q$. 
The expansion coefficients of pressure are defined by 
$p(T,\mu_q)/T^4 = \ln {\cal Z}/(VT^3) \equiv \sum_n c_n(T) (\mu_q/T)^n$,
where $V$ is spatial volume and $T$ is temperature.
Because $p$ is an analytic function of $\mu_q$ for finite volume, $c_n$ does not change even when $\mu_q$ is a complex number. Hence, the convergence radius is determined by the nearest singularity from $\mu_q=0$ in the complex $\mu_q$ plane.

In this report, we study the thermodynamic singularities of QCD 
by a simulation with relatively heavy quark mass. 
Although simulations near the chiral limit are required for the study of universality class in \cite{steph06}, it is important to study the distribution of the singularities in the complex $\mu_q$ plane as a first step toward the universality argument and for an estimation of the convergence radius even if the quark mass is not very small. 
In Sec.~\ref{sec:phase}, we discuss a method to investigate the singularities at complex $\mu_q$ and point out problems in this method.
The properties of Lee-Yang zeros are also discussed, introducing the probability distribution function of complex phase of the Boltzmann weight.  
In Sec.~\ref{sec:poten}, we estimate the distribution of the singular points using the data obtained by a simulation of two-flavor QCD with p4-improved staggered quarks of $m_{\pi}/m_{\rho} \approx 0.7$ on a $16^3 \times 4$ lattice in \cite{BS05}, and discuss the convergence radius.
Conclusions are given in Sec.~\ref{sec:conc}.

\section{ Singularities and complex phase of quark determinant with complex $\mu_q$}
\label{sec:phase}

The grand partition function at a complex chemical potential 
$\mu_q \equiv \mu_{\rm Re} +i \mu_{\rm Im}$ is given by
\begin{eqnarray}
{\cal Z}(\beta(T),\mu_{\rm Re} +i \mu_{\rm Im}) = \int 
{\cal D}U \left( \det M(\mu_{\rm Re} +i \mu_{\rm Im}) 
\right)^{N_{\rm f}} e^{-S_g}.
\label{eq:partition} 
\end{eqnarray}
To find the singularities in the complex plane, 
we define a normalized partition function,
\begin{eqnarray}
{\cal Z}_{\rm norm}(\beta, \mu_{\rm Re} +i \mu_{\rm Im} )  
\equiv 
\left| \frac{{\cal Z}(\beta, \mu_{\rm Re} +i \mu_{\rm Im} )
}{{\cal Z}(\beta,0)} \right|
= \left| \left\langle 
\left( \frac{\det M(\mu_{\rm Re} +i \mu_{\rm Im})}{\det M(0)} 
\right)^{N_{\rm f}} \right\rangle_{(\beta, \mu_q=0)} \right| ,
\label{eq:norpart} 
\end{eqnarray}
where $\left\langle \cdots \right\rangle_{(\beta, \mu_q=0)}$ means 
the expectation value at $\mu_q=0$, and $\beta=6/g^{2}$. 
Because ${\cal Z}(\beta, 0)$ is always nonzero 
for finite volume, the position of 
${\cal Z}(\beta, \mu_q)=0$ can be found by 
calculating ${\cal Z}_{\rm norm}$. 
Since ${\cal Z}$ vanishes due to the complex phase of $\det M$, i.e. 
$\theta= N_{\rm f} {\rm Im}( \ln \det M(\mu_q))$, 
we introduce the complex phase distribution function $W(\theta)$. 
The partition function is then written by 
\begin{eqnarray}
{\cal Z}_{\rm norm}(\beta,\mu_{\rm Re} +i \mu_{\rm Im}) 
= \left| \int e^{i \theta} W(\theta) d \theta \right| .
\end{eqnarray}
Identifying $\theta+2n\pi$ with $\theta$, ${\cal Z}_{\rm norm}$ vanishes when the contributions from $\theta$ and $\theta +\pi$ cancel each other.
Using the distribution function, we point out that the phase $\theta$ contains two components: 
one is related to the total quark number and produces characteristic properties of Lee-Yang zeros, and the other is irrelevant to Lee-Yang zeros and causes the sign problem at large $\mu_q$.

\begin{figure}[t]
\begin{center}
\includegraphics[width=2.5in]{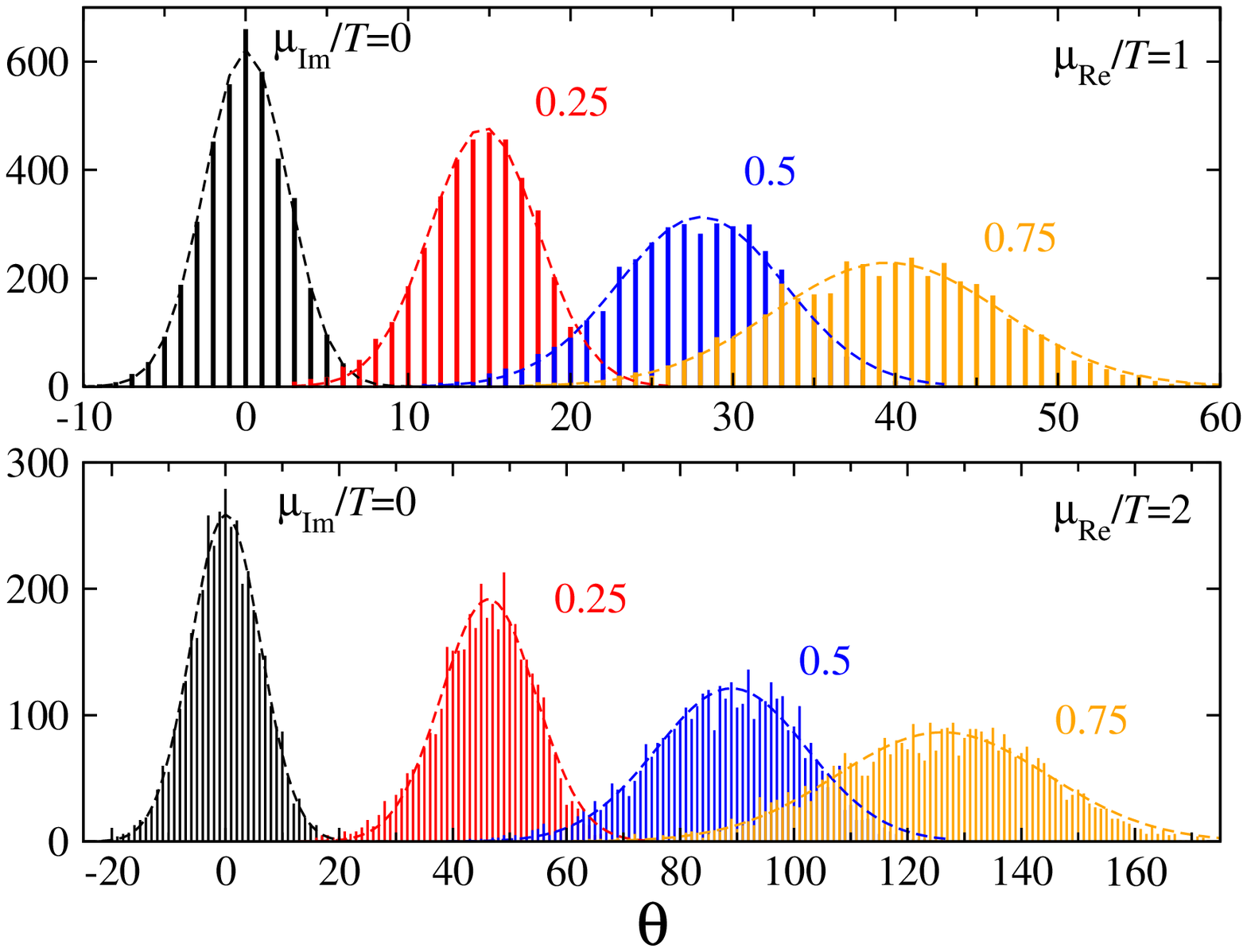}
\hskip 0.3cm
\includegraphics[width=2.5in]{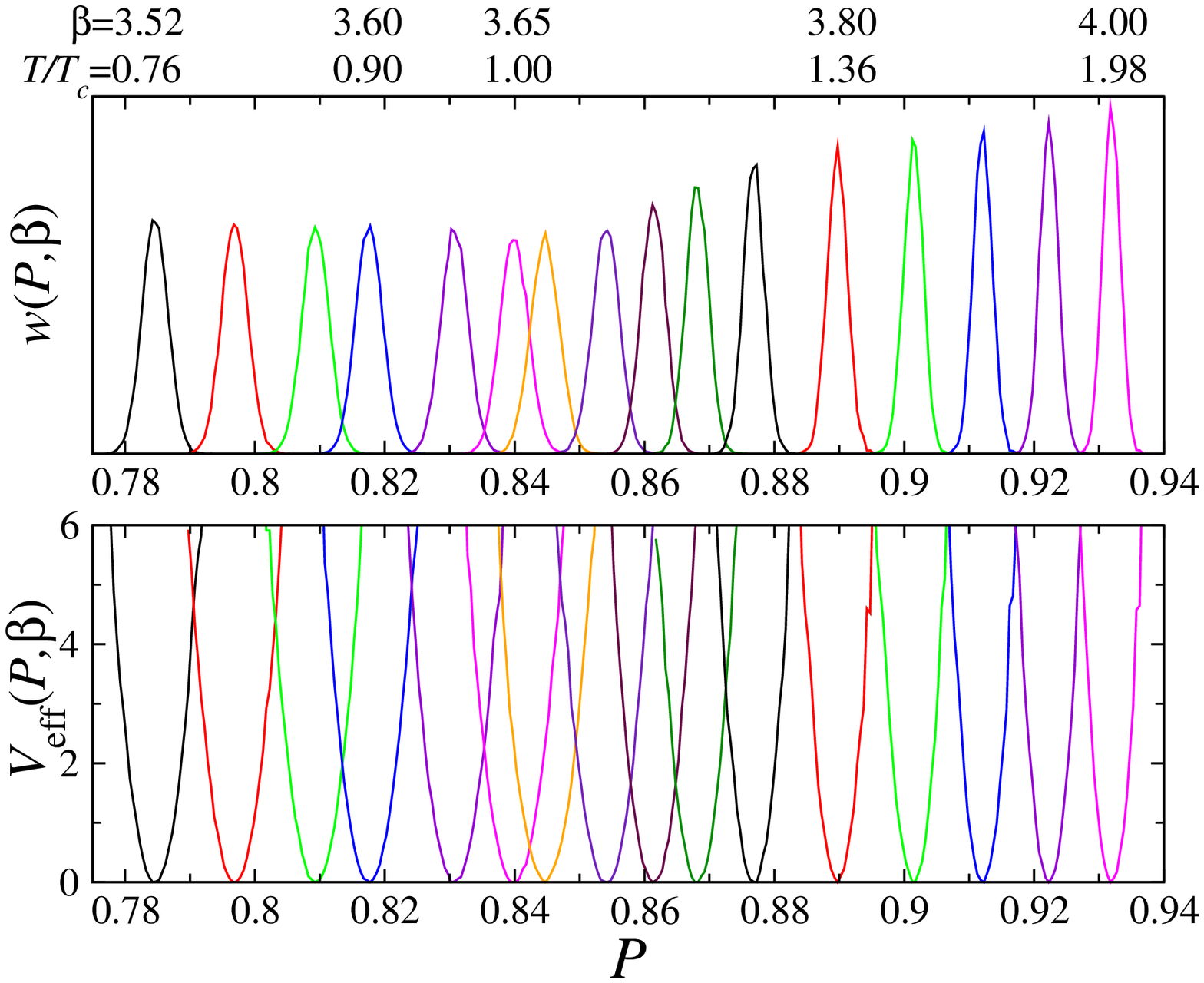}
\vskip -0.2cm
\caption{
Left: The histograms of complex phase $\theta$ for complex 
$\mu_q/T= \mu_{\rm Re}/T +i \mu_{\rm Im}/T$ at $\beta=3.65$.
The dashed curves are Gaussian functions.  
Right: The plaquette histogram $w(P,\beta)$ and the effective potential 
$V_{\rm eff}(P,\beta,0)$ 
at $\mu_q=0$ for each $\beta$.
}
\label{fig1}
\end{center}
\vskip -0.3cm
\end{figure}

We discuss the complex phase in the vicinity of the real $\mu_q$ axis. 
The phase is given by
\begin{eqnarray}
\theta (\mu_{\rm Re} +i \mu_{\rm Im}) = 
N_{\rm f} \left[ {\rm Im}( \ln \det M(\mu_{\rm Re}))
+ {\rm Re} \left( \frac{d \ln \det M(\mu_q)}{d (\mu_q/T)} 
\right)_{\mu_q=\mu_{\rm Re}} \frac{\mu_{\rm Im}}{T} + \cdots \right]. 
\label{eq:phasemuI} 
\end{eqnarray}
Because ${\cal Z}$ is real and positive for real $\mu_q$, 
the $\mu_{\rm Im}$-independent term does not contribute to ${\cal Z}=0$ for finite $V$. 
However, because of statistical fluctuations in the $\mu_{\rm Im}$-independent 
part of $\theta$, 
${\cal Z}$ may be smaller than the statistical error even at $\mu_{\rm Im}=0$. 
We plot the histogram of the complex phase in Fig.~\ref{fig1} (left).
The phase is calculated by a Taylor expansion of $\ln \det M$ up to $O(\mu_q^6)$ using data in a simulation with p4-improved staggered quarks \cite{BS05}.
The result of $\mu_{\rm Im}/T=0$ is given by the $\mu_{\rm Im}$-independent part only. 
The magnitude of ${\cal Z}$ decreases exponentially as $\mu_{\rm Re}$ increases, 
i.e. ${\cal Z} \sim \exp[- \langle \theta^2 \rangle /2]$ 
with $\langle \theta^2 \rangle \sim O(\mu_{\rm Re}^2)$ 
for $\mu_{\rm Im}=0$ \cite{eji07}. 
Once ${\cal Z}$ becomes smaller than the statistical error at large $\mu_q$, 
${\cal Z}_{\rm norm}$ vanishes at random.
Such ${\cal Z}_{\rm norm}=0$ are irrelevant to Lee-Yang zeros, 
but one cannot distinguish such fake zeros from real ones. 
This is a kind of the sign problem.


Another interesting point is that the operator $d (\ln \det M(\mu_q))/d (\mu_q/T)$ 
in the second term of Eq.~(\ref{eq:phasemuI}) corresponds to the quark number 
$(N)$ on each configuration. 
Therefore, the distribution of the complex phase is related to 
the distribution of the quark number at the leading order of $\mu_{\rm Im}$.
Let us consider a canonical ensemble with fixed $N$.
The canonical partition function ${\cal Z}_C(T,N)$ for each $N$ 
is related to the grand partition function ${\cal Z}(T, \mu_q)$  
through a fugacity expansion, 
\begin{eqnarray}
{\cal Z}(T,\mu_q) 
= \sum_{N} \ {\cal Z}_C(T,N) e^{N \mu_q/T}
\equiv \sum_{N} \ W(T,N) e^{i \bar{\theta}} 
= \int V W(T, V \rho) e^{i \rho V \mu_{\rm Im}/T} d \rho,
\label{eq:cpartition} 
\end{eqnarray}
where $\rho$ is the quark number density.
In the case of 
$\mu_q = \mu_{\rm Re} +i \mu_{\rm Im}$,
the complex phase in ${\cal Z}(T,\mu_q)$ is 
$\bar{\theta} = N \mu_{\rm Im}/T = \rho V \mu_{\rm Im}/T$, 
since ${\cal Z}_C(T,N)$ is real and positive. 
$W(T,N)$ is the distribution function of the total quark number, 
$W(T,N)={\cal Z}_C(T,N) e^{N \mu_{\rm Re}/T}$. 

The $\mu_{\rm Im}$-independent part in this $\bar{\theta}$ is eliminated if the canonical partition function is obtained by a partial path integral with fixed $N$. 
Once the problem of the unnecessary part of the phase is solved, one can discuss the order of phase transitions in the following way.

We find from Eq.~(\ref{eq:cpartition}) that 
${\cal Z}$ as a function of $(V \mu_{\rm Im}/T)$ is obtained through 
a Fourier transformation of $W$ with respect to $\rho$. 
In the case of a normal point of $\mu_{\rm Re}$ or 
a crossover pseudo-critical point, where the distribution 
is expected to be of Gaussian for sufficiently large $V$, 
${\cal Z}$ does not vanish except in the limit of
$V \mu_{\rm Im}/T \to \pm \infty$ because the 
function which is obtained through a Fourier transformation of a Gaussian 
function again is a Gaussian function. 
Therefore, the complex $\mu_q$ at which ${\cal Z}=0$ does not go to 
the real axis in the infinite volume limit. 

On the other hand, at a first order phase transition point, 
two phases having a different quark number coexist.
In this case, we expect that $W(T,V \rho)$ has two peaks having 
the same peak height at the transition point.
Performing the Fourier transformation of such a double peaked function 
leads to a function which has zeros periodically. 
For example, a distribution function $W(T, V \rho)$ having two Gaussian 
peaks at $\rho_1$ and $\rho_2$ leads to ${\cal Z}$ which has zeros at 
$ \mu_{\rm Im}/T = (2n+1) \pi / [V(\rho_2 -\rho_1)],$
with $n=0,1,2,3,\cdots$.
The Lee-Yang zeros approach the real axis as $1/V$.  
Therefore, the study of $1/V$ scaling of the Lee-Yang zero is equivalent to finding 
the double-peak structure at a first order phase transition.
The same discussion for SU(3) pure gauge theory is given in \cite{eji05}.

\section{Grand canonical partition function by a density of state method}
\label{sec:poten}

As noted above, the numerical study in terms of the singularities of ${\cal Z}=0$ has a potential danger for large $\mu_q$. 
However, 
the position of ${\cal Z}=0$ can be estimated from the distribution function of the complex phase, since ${\cal Z}$ vanishes when the distribution function has two peaks or more and the contributions from these peaks cancel each other.
In this study, we investigate the distribution function of the complex phase instead of ${\cal Z}$. 
Once the quark number is fixed, the distribution function is related to 
the canonical partition function, as discussed in \cite{eji08} for real $\mu_q$. 
However, fixing the quark number is not essential for eliminating the sign problem,
and the calculation of ${\cal Z}_C (T,N)$ is not easy actually. 
We rather use a density of state method with fixing the plaquette variable 
and apply an approximation proposed in \cite{eji07} to avoid the sign problem. 

We introduce a probability distribution function 
of the plaquette, which is defined by 
\begin{eqnarray}
W(P',\beta,\mu_q) = \frac{1}{{\cal Z} (\beta,0)} 
\int {\cal D} U \ \delta(P'-P) \ (\det M)^{N_{\rm f}} 
e^{6\beta N_{\rm site} P}, 
\label{eq:pdist}
\end{eqnarray}
where $\delta(x)$ is the delta function. 
For later discussions, we define the average plaquette $P$ as 
$P \equiv -S_g/(6 \beta N_{\rm site})$ 
and the quark matrix $M$ as independent of $\beta$. 
$N_{\rm site}$ is the number of sites.
The plaquette distribution functions for $\mu_q=0$ are shown in 
Fig.~\ref{fig1} (right) for each $\beta$. 
We denote $ w(P,\beta) \equiv W(P,\beta,0)$.
The normalized partition function is rewritten as 
\begin{eqnarray}
\frac{{\cal Z}(\beta, \mu_q)}{{\cal Z}(\beta, 0)} 
= \int W(P,\beta,\mu_q) \ dP
= \int R(P,\mu_q) w(P,\beta) \ dP.
\label{eq:rewmu}
\end{eqnarray}
Here, $R(P,\mu_q)$ is the reweighting factor for finite $\mu_q$ 
defined by 
\begin{eqnarray}
R(P',\mu_q) \equiv 
\frac{\int {\cal D} U \ \delta(P'-P) (\det M(\mu_q))^{N_{\rm f}}}{
\int {\cal D} U \ \delta(P'-P) (\det M(0))^{N_{\rm f}}} 
= \frac{ \left\langle \delta(P'-P) 
\frac{(\det M(\mu_q))^{N_{\rm f}}}{(\det M(0))^{N_{\rm f}}} 
\right\rangle_{(\beta, \mu_q=0)} }{
\left\langle \delta(P'-P) \right\rangle_{(\beta, \mu_q=0)}}.
\label{eq:rmudef}
\end{eqnarray}
This $R(P, \mu_q)$ is independent of $\beta$, and 
$R(P, \mu_q)$ can be measured at any $\beta$. In this method, 
all simulations are performed at $\mu_q=0$ and the effect of 
finite $\mu_q$ is introduced though the operator 
$\det M(\mu_q) / \det M(0)$ measured on the configurations 
generated by the simulations at $\mu_q=0$.

Because QCD has time-reflection symmetry, the partition function is invariant 
under a change from $\mu_q$ to $-\mu_q$, i.e. $R(P,-\mu_q) = R(P,\mu_q)$.
Moreover, the quark determinant satisfies $\det M (-\mu_q) = (\det M(\mu_q^*))^*$.
From these equations, we get 
\begin{eqnarray}
[R(P, \mu_q)]^* = R(P, \mu_q^*).
\end{eqnarray}
This indicates that $R(P, \mu_q)$ is real in the case of real $\mu_q$, 
i.e. $\mu_q = \mu_q^*$.
Then, the probability distribution of the plaquette given by 
$R(P, \mu_q) w(P, \beta)$ is real. 
However, once the imaginary part of $\mu_q$ becomes nonzero, 
$R(P, \mu_q)$ is not a real number any more.
We thus write the partition function,
\begin{eqnarray}
\frac{{\cal Z}(\beta, \mu_q)}{{\cal Z}(\beta, 0)} 
= \int e^{i \phi(P,\mu_q)} |R(P,\mu_q)| w(P,\beta) \ dP
= \int e^{i \phi(P,\mu_q)} |R(P,\mu_q)| w(P,\beta) 
\left( \frac{d \phi}{d P} \right)^{-1} d\phi.
\label{eq:phidis}
\end{eqnarray}
Because this complex phase $\phi$ vanishes at $\mu_{\rm Im}=0$, 
$\phi$ does not have the $\mu_{\rm Im}$-independent part,  
and the problem of the statistical error due to the 
$\mu_{\rm Im}$-independent part is eliminated. 
As we discussed in the previous section, 
if $|R(P,\mu_q)| w(P,\beta) (d \phi /d P)^{-1}$ becomes a 
double-peaked function of $\phi$ and the distance between these peaks is 
equal to $(2n+1)\pi$ with an integer $n$, 
the partition function becomes zero, i.e. a Lee-Yang zero appears.
To investigate whether the probability distribution has double-peak or not, 
we introduce an effective potential defined by 
\begin{eqnarray}
V_{\rm eff}(\phi, \beta, \mu_q) & \equiv & 
-\ln \left| R(P(\phi), \mu_q) \right| - \ln w(P(\phi), \beta) 
+ \ln \left( \frac{d \phi}{d P} \right)_{\rm norm} (P(\phi), \mu_q). 
\label{eq:potmu}
\end{eqnarray}
Here, we normalize the value of $d \phi/dP$ by the maximum value for each $\mu_q$. 
This effective potential is a function of $P$ in practice. 
We define 
$V_{\rm eff}(P, \beta, \mu_q) \equiv V_{\rm eff}(\phi(P), \beta, \mu_q)$.

Performing Monte-Carlo simulations, we calculate these three quantities, $|R(P,\mu_q)|$, $\phi (P,\mu_q)$ and $w(P,\beta)$.
However, the exact calculation of the quark determinant is difficult except on small lattices. 
In this study, we estimate the quark determinant from the data of Taylor expansion coefficients up to $O(\mu_q^6)$ around $\mu_q=0$ obtained by a simulation of two-flavor QCD with p4-improved staggered quarks in \cite{BS05}.
The truncation error has been discussed in \cite{eji07}.
Because we define $\theta$ by the Taylor expansion of $\ln \det M$, $\theta$ is not restricted to the range from $-\pi$ to $\pi$.

Next, we discuss the sign problem in the calculation of $V_{\rm eff}$.
We denote the quark determinant as
$N_{\rm f} \ln[\det M(\mu_q) / \det M(0)] \equiv F +i \theta$.
Histograms of $\theta$ are shown in Fig.~\ref{fig1} (left). 
We fitted the histograms to Gaussian functions. The results are the dashed curves. The distributions seem to be well-approximated by Gaussian functions.
Here, we perform a cumulant expansion, 
\begin{eqnarray}
\langle \exp(F +i \theta) \rangle = \langle e^F \rangle 
\exp \left[ i\langle \theta \rangle 
- \frac{1}{2} \langle (\Delta \theta)^2 \rangle
- \frac{i}{3!} \langle (\Delta \theta)^3 \rangle 
+ i \langle \Delta F \Delta \theta \rangle 
- \frac{1}{2} \langle \Delta F (\Delta \theta)^2 \rangle + \cdots \right]
\label{eq:cum}
\end{eqnarray}
with $\Delta X= X - \langle X \rangle$. 
If the distribution of $\theta$ is of Gaussian, the $O(\theta^n)$ terms vanish for $n >2$ in this equation \cite{eji07}. 
Moreover, since $\theta \sim O(\mu_q)$ and $F \sim O(\mu_q^2)$, this expansion can be regarded as a power expansion in $\mu_q$. 
If the expansion of Eq.~(\ref{eq:cum}) is good, we can extract the phase factor $e^{i \phi}$ from $R$ easily and the sign problem in $|R|$ is eliminated.
We deal with the first two terms, 
i.e. $i \langle \theta \rangle$ and 
$-\langle (\Delta \theta)^2 \rangle /2$,
assuming the Gaussian distribution. 
The correlation terms between $F$ and $\theta$ are also neglected as a first step.
Because we calculate the expectation value with fixed $P$ and the values of $F$ and $P$ are strongly correlated, the $\Delta F$ may be small once $P$ is fixed.
Then, $\phi \approx \langle \theta \rangle$.

\begin{figure}[t]
\begin{center}
\includegraphics[width=1.95in]{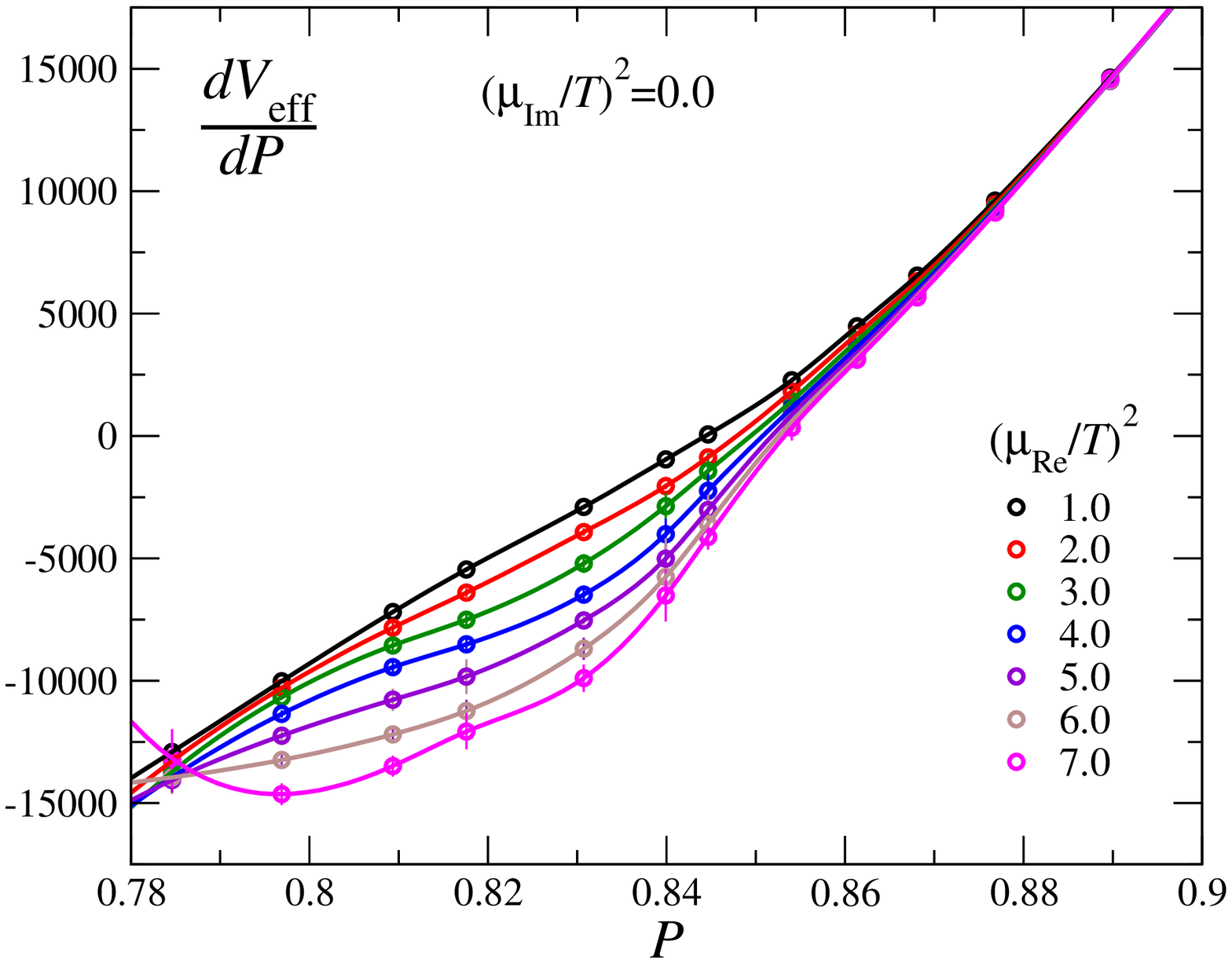}
\includegraphics[width=1.95in]{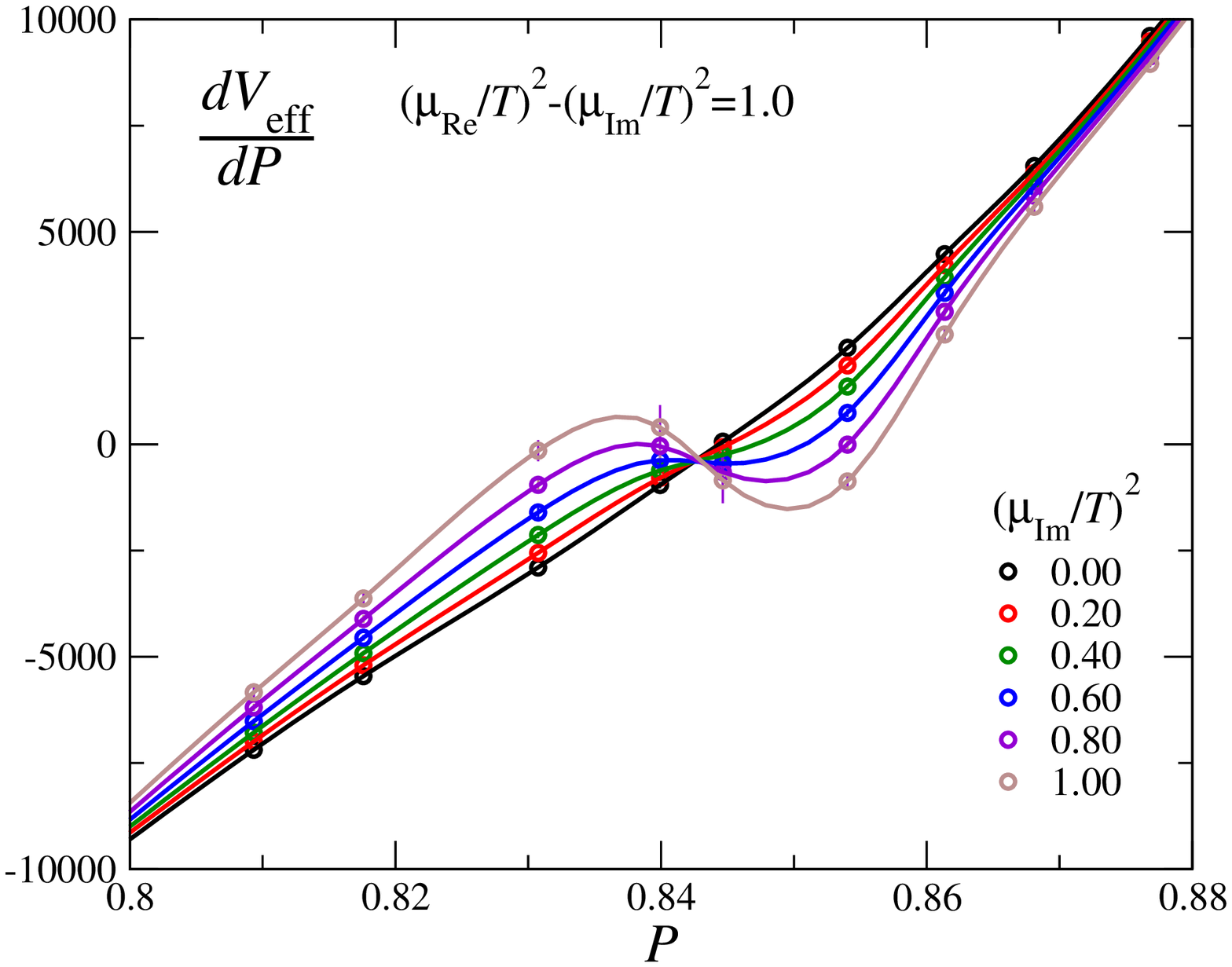}
\includegraphics[width=1.95in]{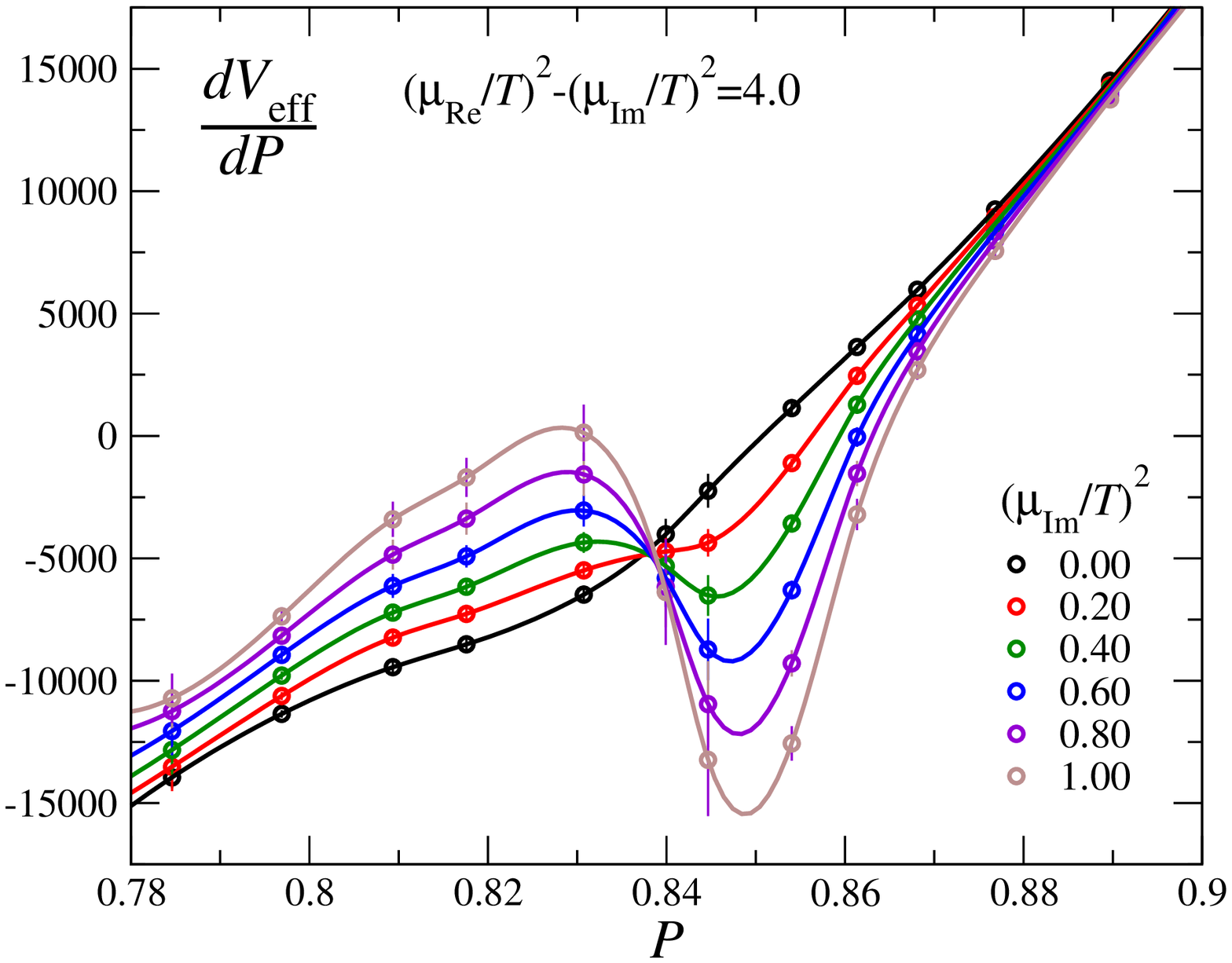}
\vskip -0.2cm
\caption{
The derivative of the effective potential $dV_{\rm eff}/dP$ at $\beta=3.65$ 
for $(\mu_{\rm Im}/T)^2 =0.0$ (left),
$(\mu_{\rm Re}/T)^2 - (\mu_{\rm Im}/T)^2 =1.0$ (middle) and $4.0$ (right). 
We measured them at the peaks of the plaquette histograms in Fig.~{\protect \ref{fig1}} (right) and interpolated the data by a cubic spline method.}
\label{fig2}
\end{center}
\vskip -0.3cm
\end{figure} 

The results of the first derivative of $V_{\rm eff}(P,\beta,\mu_q)$ are shown in Fig.~\ref{fig2} for various $\mu_q/T$ with $(\mu_{\rm Im}/T)^2=0$ (left) and with ${\rm Re} [(\mu_q/T)^2]= (\mu_{\rm Re}/T)^2-(\mu_{\rm Im}/T)^2=1$ (middle) and $4$ (right). $\beta=3.65$ is adopted for these results.
We discuss $d V_{\rm eff}/dP$ instead of $V_{\rm eff}$ itself because $d^2 V_{\rm eff}/dP^2$ is independent of $\beta$. 
$d V_{\rm eff}/dP$ at different $\beta$ can be estimated by the equation,  
\begin{eqnarray}
\frac{d V_{\rm eff}}{dP} (P,\beta)
= \frac{d V_{\rm eff}}{dP} (P,\beta_0) -6 (\beta - \beta_0) N_{\rm site},
\label{eq:derrewbeta}
\end{eqnarray}
under the parameter change from $\beta_0$ to $\beta$. 
This equation is derived from the equation, 
$ w(P, \beta) = e^{6 (\beta - \beta_0) N_{\rm site} P} w(P, \beta_0), $
which is given from the definition, Eq.~(\ref{eq:pdist}). 
Therefore, the form of $dV_{\rm eff}/dP$ as a function of $P$ does not change 
with $\beta$ up to a $\beta$-dependent constant.
Using the behaviors of $dV_{\rm eff}/dP$, the value of $P$ minimizing $V_{\rm eff}$, 
i.e. $d V_{\rm eff}/dP=0$, can be also controlled by $\beta$.

If the effective potential $V_{\rm eff}$ is a double-well function of $\phi$, 
$d V_{\rm eff}/d \phi$ is an S-shaped function and vanishes three times.
In such a case, there exists a region of $P$ where the derivative of $ d V_{\rm eff}/d P = (d V_{\rm eff}/d\phi) (d\phi/dP)$ is negative, since $\phi$ is a monotonically increasing function of $P$. 
The left panel of Fig.~\ref{fig2} shows that the region of $d^2 V_{\rm eff}/d P^2 <0$ appears at high density, i.e. $(\mu_{\rm Re}/T)^2 > 6$ for $\mu_{\rm Im}/T=0$. Also, in the region of large $\mu_{\rm Im}/T$, $d V_{\rm eff}/d P$ becomes an S-shaped function, which is shown in the middle and right panels of Fig.~\ref{fig2} for $(\mu_{\rm Re}/T)^2 = 1$ and $4$. 

\begin{figure}[t]
\begin{center}
\includegraphics[width=2.5in]{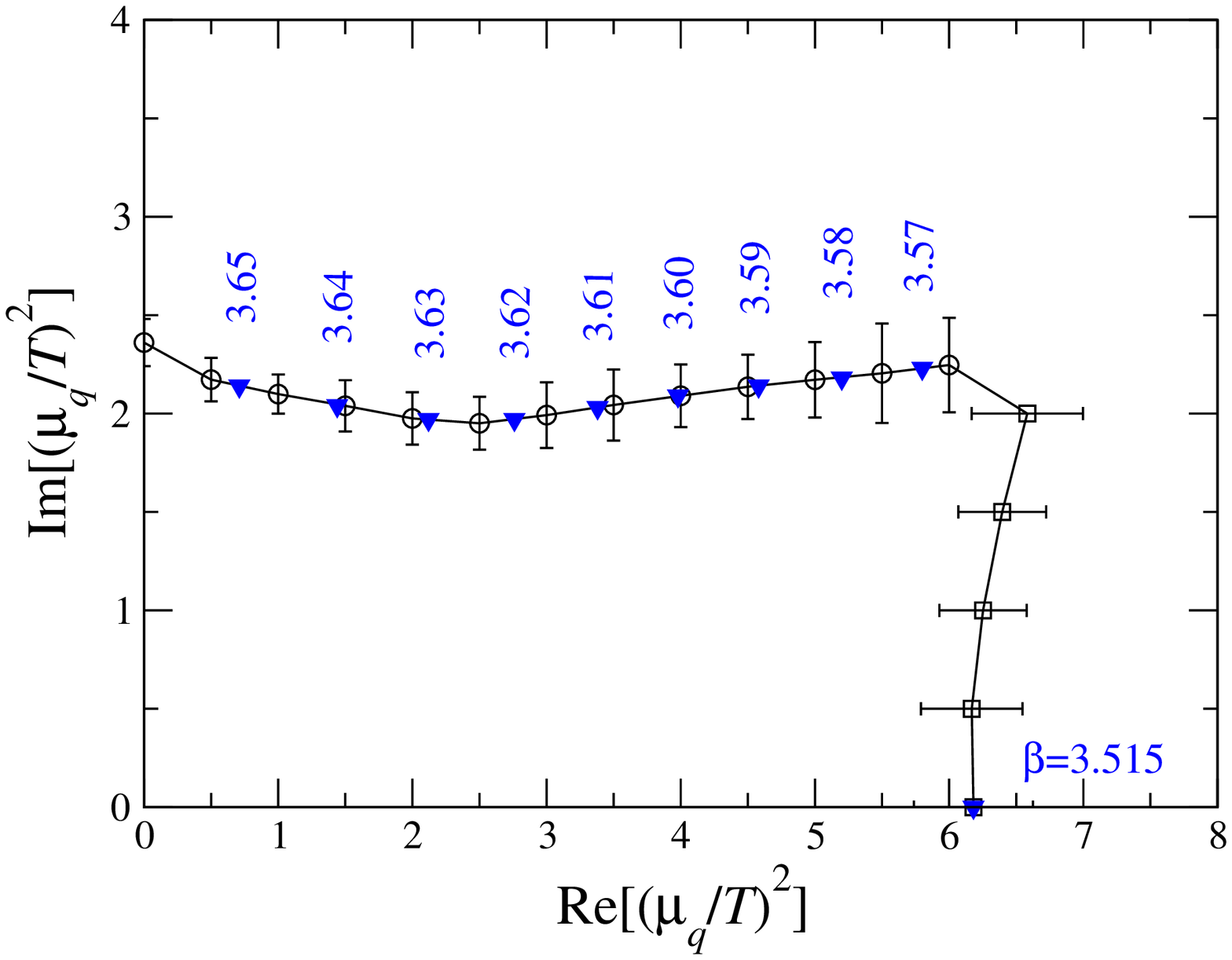}
\hskip 0.5cm
\includegraphics[width=2.5in]{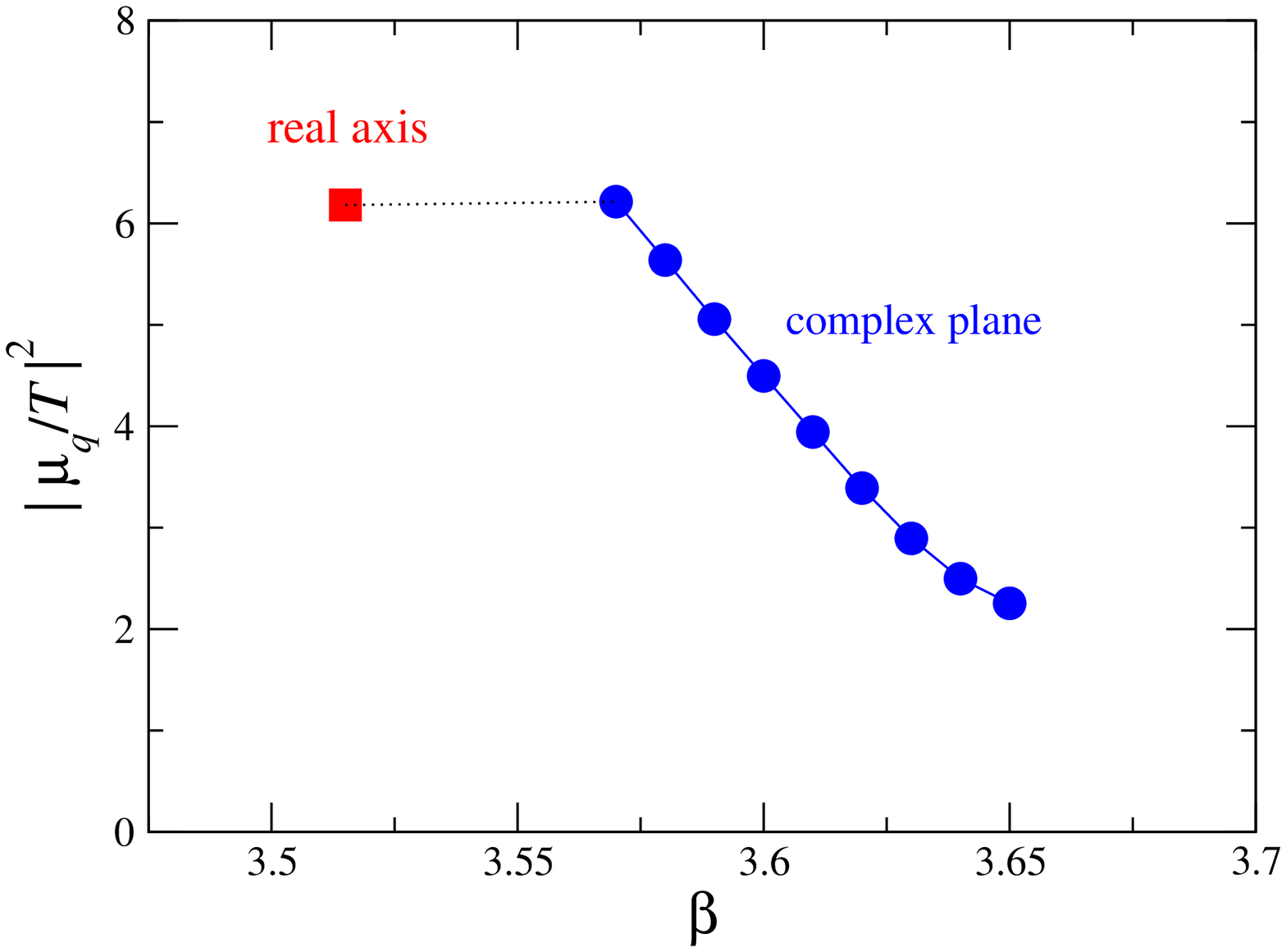}
\vskip -0.2cm
\caption{
Left: Boundary where $V_{\rm eff}$ changes to double-well type 
in the complex $(\mu_q/T)^2$ plane.
Below this line, $V_{\rm eff}$ is always of single-well.
The numbers in this figure are the values of $\beta$ at which 
$dV_{\rm eff}/dP$ and $d^2 V_{\rm eff}/dP^2$ vanish simultaneously.
Right: The distance to the boundary form $\mu_q=0$ for each $\beta$. 
This corresponds to the radius of convergence.
The red symbol is the results on the real axis.
}
\label{fig3}
\end{center}
\vskip -0.3cm
\end{figure}

Next, we investigate the boundary at which $dV_{\rm eff}/d P$ changes to an S-shaped function from a monotonic function, which is shown as a line in Fig.~\ref{fig3} (left).
On this line, $d^2 V_{\rm eff}/dP^2=0$ at a value of $P$. 
The constant part of $dV_{\rm eff}/d P$ is changed by $\beta$.
The values of $\beta$ at which both $d^2 V_{\rm eff}/dP^2$ and $d V_{\rm eff}/dP$ vanish simultaneously are indicated in this figure. 
Above this line (large ${\rm Im}(\mu_q^2)$ or large ${\rm Re}(\mu_q^2)$) with this $\beta$, the double-well potential appears, and the phase cancelation occurs near this line.
The temperatures corresponding to these values of $\beta$ are shown above the curves of $w(P,\beta)$ in Fig.~\ref{fig1} (right).
This result suggests the existence of singularities (Lee-Yang zeros) in the region of large ${\rm Im}(\mu_q^2)$ as well as the region of large ${\rm Re}(\mu_q^2)$, 
and the boundary at large ${\rm Im}(\mu_q^2)$ is closer to $\mu_q=0$ than that at large ${\rm Re}(\mu_q^2)$.
The distance to the boundary from $\mu_q=0$ is shown in Fig.~\ref{fig3} (right) for each $\beta$. 
The blue circles are the distance to the line in the complex plane and the red square is that on the real axis.
The boundary on the real axis is essentially the same as the critical point in \cite{eji07}. 
This distance to the boundary is approximately equal to the convergence radius 
of an expansion of $\ln {\cal Z}/(VT^3) = \sum_n c_{2n} (\mu_q/T)^{2n}$.

\section{Conclusions}
\label{sec:conc}
We studied the singularities of QCD in the complex $\mu_q$ plane by a numerical simulation. 
Because the sign problem makes the calculation of the partition function difficult,
we discussed a probability distribution function of a complex phase instead of the partition function itself.
In this calculation, we used a kind of reweighting method together with 
the approximation in \cite{eji07}:
the quark determinant is estimated by a Taylor expansion, and the Gaussian distribution of the complex phase of $\det M$ is assumed. 
We found that there is a region where the phase distribution has two peaks, suggesting the existence of singularities, in the region of large ${\rm Im}( \mu_q^2)$ as well as of large ${\rm Re}(\mu_q^2)$. 
We moreover estimated the distance to the nearest singularity from $\mu_q=0$ in terms of $\mu_q^2$ for each temperature. 
The distance is regarded as the convergence radius of a Taylor expansion of the thermodynamic potential.
Although our simulation is performed with relatively heavy quark mass, the result suggests that convergence radius at the temperature of the real QCD critical point $(T_{cp})$ may be longer than the convergence radius in the crossover region at $T>T_{cp}$.


\end{document}